\documentclass{article}
\usepackage[utf8]{inputenc}
\usepackage{authblk}
\usepackage[dvips]{graphicx}
\graphicspath{{noiseimages/}}
\usepackage{xcolor}
\usepackage[T2A]{fontenc}
\usepackage{tikz}
\usepackage{pgfplots}
\usepackage{booktabs}
\usepackage{mathrsfs}
\pgfplotsset{compat=1.7}
\usetikzlibrary{intersections, pgfplots.fillbetween}
\usetikzlibrary{decorations.pathmorphing}
\usetikzlibrary{arrows.meta}
\usepackage[mode=buildnew]{standalone}
\usepackage[toc,page]{appendix}
\usepackage{amsmath}
\usepackage{amssymb}
\usepackage{float}
\usepackage{comment}
\usepackage{a4wide}
\usepackage[english]{babel}
\usepackage{hyperref}
\definecolor{urlcolor}{HTML}{990000}
\definecolor{linkcolor}{HTML}{005F5F}
\hypersetup{pdfstartview=FitH,  linkcolor=linkcolor,urlcolor=urlcolor, colorlinks=true,citecolor=blue}
\usepackage[page]{appendix}

\author[1,2]{D. V. Diakonov\footnote{\texttt{dmitrii.dyakonov@phystech.edu}}}

\affil[1]{Institutskii per. 9, Moscow Institute of Physics and Technology, 141700, Dolgoprudny, Russia}

\affil[2]{\itshape Bol'shoi Karetnyi per., 19, Institute for Information Transmission Problems, 127994, Moscow, Russia}

\title{\textcolor{black}{One more Sine-Gordon soliton in AdS}}

\begin{document}

\numberwithin{equation}{section}

\maketitle

\begin{abstract}
In previous work \cite{Akhmedov:2026msi}, we found soliton solutions in a deformation of the sine-Gordon theory in AdS spacetime that, in the infinite-radius limit, reduce to single-soliton solutions in flat space. In this paper, we find another single soliton solution that has no analog in flat space.
\end{abstract}

\section{Introduction}

Soliton solutions of nonlinear equations are important in classical and quantum field theory. We ask how such solitons extend to maximally symmetric curved spaces (dS, AdS, Lobachevsky) with fixed background. Since QFT beyond tree level in dS/AdS is difficult, a simple integrable non-conformal model without large N is desirable; two-dimensional sine-Gordon theory is a promising candidate. The question is whether it is integrable in hyperbolic spaces and in what sense. In flat space, integrability is tied to infinitely many integrals of motion and soliton solutions via inverse scattering; in curved space conservation is subtle. We aim to determine whether sine-Gordon solitons exist in maximally symmetric space. In previous work \cite{Akhmedov:2026msi} we found that in the deformation of the sine-Gordon theory in $AdS_{d+1}$ ($dS_{d+1}$ and  $H_d$):
\begin{align}
    \Box \phi - m^2 \sin \phi - \frac{2 d m}{R} \sin \frac{\phi}{2}=0, 
\end{align}
where $R$ is the radius of the corresponding symmetric hyperbolic space and $\Box$ is the curved-space d'Alembert operator, one can construct only single soliton solutions:
\begin{itemize}
\item In $(1+1)$-dimensional $AdS$:
\begin{align}
\label{one solitons 1}
    \phi= 4 \arctan\left[  \left(\frac{X \cdot \xi }{R}\right)^{m R}\right],
\end{align}
where $mR\in \mathbb{N}^+$. In this case there is only a single linearly independent lightlike vector $(\xi \cdot \xi)=0$ in the ambient spacetime.
\item In $(d+1)$-dimensional $AdS$ spacetime with $d \ge 2$, the solution \eqref{one solitons 1} can be generalized to:
\begin{align}
    \phi= 4 \arctan \left[ \left(\frac{X \cdot \xi}{R}\right )^{mR} F\left(\frac{(X \cdot\xi_{i_1}) }{(X \cdot\xi_{j_1})},...,\frac{(X \cdot\xi_{i_p}) }{(X \cdot\xi_{j_p})}\right) \right],
\end{align}
where $X$ denotes the ambient spacetime coordinates and $\xi_i$, $i=1,\dots,N$, is a set of mutually orthogonal null vectors from a two-dimensional null vector space, $(\xi_i \cdot \xi_i)=0$ and $(\xi_i \cdot \xi_j)=0$ for $i \neq j$.  In this case the function $F$ defines the solitonic profile in the "transverse" direction.
\end{itemize}
Note also that this theory reduces in the limit $m/R \to 0$ to the standard sine-Gordon theory. The extra terms in the equation are actually quite natural, because they arise from the coupling of the field to the curvature of spacetime and are not merely a modification of the potential.

However, several questions remain:
\begin{itemize}
    \item Is it possible to construct multi-soliton solutions in $1+1$ dimensions using hyperbolic plane waves $ (X \cdot \xi_i )^{m R}$ with null vectors that are not orthogonal, $(\xi_i \cdot \xi_i)=0$ and $(\xi_i \cdot \xi_j) \ne 0$ for $i \neq j$? Is the theory integrable, and in what sense?
    \item Is there a relation between the extra term in the potential and the extra term that appears in the supersymmetric generalization of the sine-Gordon theory \cite{Inami:1995np}, with action:
\begin{align}
    S=\int d^2 x \left(\frac{1}{2}\partial_\mu \phi \partial^\mu \phi-i\bar{\psi} \gamma^\mu \partial_\mu \psi +m^2 \cos \left(\phi\right) +2 m \bar{\psi} \psi \cos\left( \frac{\phi}{2}\right) \right).
\end{align}
In fact, if in curved spacetime the theory dynamically generates an expectation value of the operator $\bar{\psi} \psi$ might be proportional to the curvature, then for the scalar field one obtains the above equation.
\item Is it possible to generalize the approach to studying soliton-like solutions to other nonlinear models? Thus, in a recent paper \cite{Sadekov:2026gmc}, it was shown that, using the embedding-space formalism with an auxiliary null vector, the author derived first-order (Bäcklund-like) equations whose integrability yields the solution of the Liouville equation in a $(d+1)$-dimensional maximally symmetric spaces.
\item Is it possible to construct soliton solutions with non-null vectors $(\xi \cdot \xi)\ne0$?
\end{itemize}

In the present paper we answer the last question and show that in the deformed sine-Gordon theory with another coefficient in front of $\sin(\phi/2)$:
\begin{align}
    \Box \phi - m^2 \sin \phi - 2 \left(m^2+\frac{d-1}{R^2}\right) \sin \frac{\phi}{2} = 0,
\end{align}
there is a single-soliton solution
\begin{align}
     \phi = 4 \arctan \left( \frac{\xi \cdot X }{R}\right),
\end{align}
where the constant vector $\xi$ is non-null and satisfies the normalization condition:
\begin{align}
    (\xi \cdot \xi) = 1 - (m R)^2.
\end{align}
This solution is new, and does not have a flat-space analog, since it reduces to a constant solution in the flat-space limit: $\phi \to 2\pi$.

We also note that in $dS_{d+1}$ and in Lobachevsky space one can similarly construct such solutions; this is a simple generalization to other maximally symmetric spaces. We also mention that in $AdS_{d+1}$ one can consider a deformation of the $\phi^4$ potential in order to construct a kink solution using a non-null vector $\xi$. This solution is analogous to the null-vector solution and has a form: $\phi=\tanh \log (X\cdot \xi)$, but it corresponds to a quarter potential with two minima, one of which is global; a detailed discussion of this problem is beyond the scope of the present paper.

\section{Geometry of hyperbolic spaces and hyperbolic plane waves}

A $(d+1)$-dimensional maximally symmetric space can be embedded in a flat spacetime of $d+2$ dimensions. $AdS$ spacetime is the hyperboloid embedded in a $(d+2)$-dimensional ambient flat spacetime with signature $(-,-,+,...,+)$:
\begin{align}
AdS_{d+1}=\{ X  \in \mathbf{R}^{2,d}, \ (X\cdot X)=X_\alpha X^\alpha =-R^2 \}, \quad \alpha = \overline{1,d+2}.
\end{align}
In what follows, we will set the radii of these hyperboloids to $R = 1$, unless otherwise stated.

Now let us introduce the hyperbolic plane waves:
\begin{align}
    f(X) = (X\cdot \xi),
\end{align}
which solve the Klein-Gordon equation:
\begin{align}
\label{box12}
   \Box (X \cdot \xi)=(1+ d)  (X \cdot\xi),
\end{align}
and satisfy the property:
\begin{align}
    \nabla_\mu (X\cdot \xi_1 ) \nabla^\mu (X \cdot \xi_2) =  (X\cdot \xi_1 ) (X \cdot \xi_2)+ (\xi_1 \cdot \xi_2).
    \label{waveprop}
\end{align}
For more detail see \cite{Akhmedov:2026msi}. This property holds for any maximally symmetric space \cite{Bros:1994dn,Bros:1995js,Moschella:2025lqy,Moschella:2007zza,Akhmedova:2019bau}.

\section{New solution}
\begin{figure}[H]
    \centering
    \includegraphics[scale=0.24]{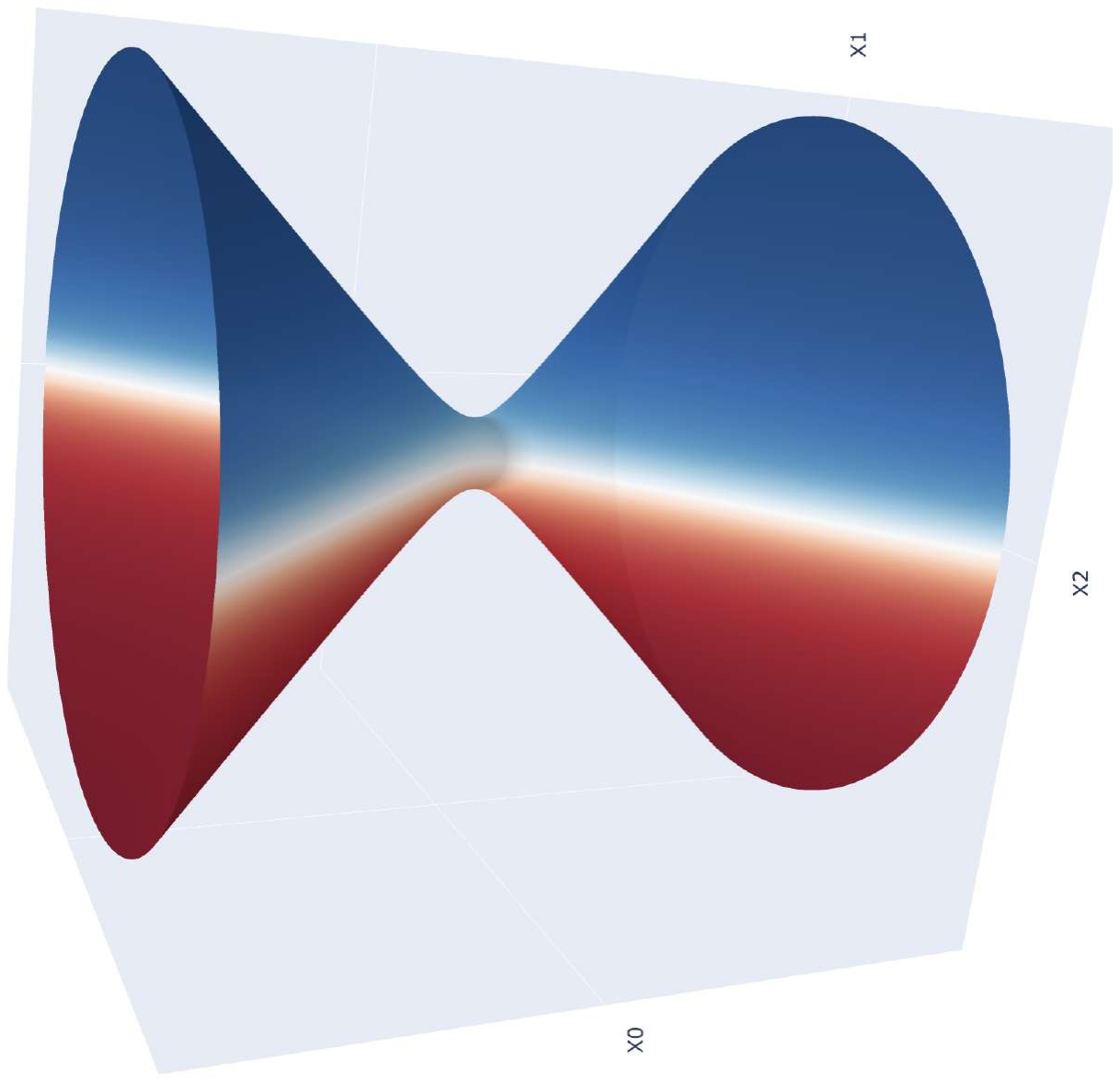}
       \includegraphics[scale=0.24]{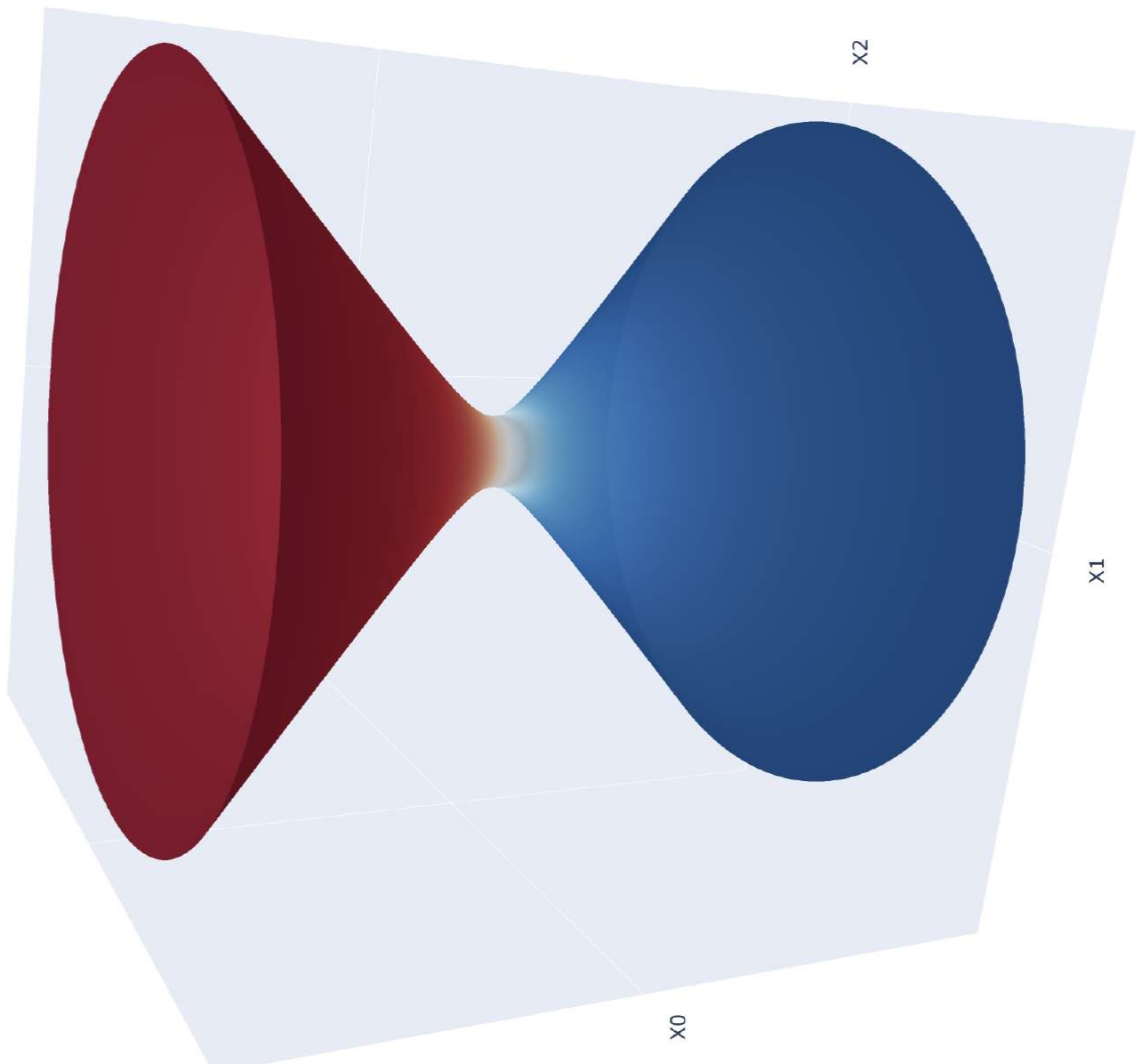}
    \caption{The value of the $\phi$ field of the soliton in $AdS$ for $(\xi\cdot \xi)>0$ and $(\xi\cdot \xi)<0$. The field value changes from $-2\pi$ (red) to $2\pi$ (blue).}
    \label{Glob de sitter 1}
\end{figure}

Consider a $(d+1)$-dimensional AdS spacetime. The double sine-Gordon equation is:
\begin{align}
\label{double sin gor}
    \Box \phi - m^2 \sin \phi - 2 q \sin \frac{\phi}{2} = 0,
\end{align} 
where $q$ is not fixed.
To find nontrivial solutions, let us use the ansatz:
\begin{align}
    \phi = 4 \arctan(G).
\end{align}
Substituting this ansatz into Eq.~\eqref{double sin gor}, one can rewrite the double sine-Gordon equation in the form:
\begin{gather}
\label{two line}
   \left( \Box -  (m^2+q) \right) G \\
   + G \left( G \Box G - 2 \nabla_\mu G \nabla^\mu G + (m^2-q) G^2 \right) = 0. \nonumber
\end{gather}
Let
\begin{align}
    G = \xi \cdot X,
\end{align}
where $\xi$ is a constant non-null vector. Using \eqref{box12} and \eqref{waveprop}, the equation yields:
\begin{gather}
   \left( d+1 - (m^2+q) \right) G \\
   + G \left( (d+1) G^2 - 2 \bigl(G^2 + (\xi \cdot \xi)\bigr) + (m^2-q) G^2 \right) = 0. \nonumber
\end{gather}
After algebraic simplification we obtain:
\begin{gather}
   \left( d+1 - (m^2+q) -2 (\xi \cdot \xi)\right) G \\
   + \left( d-1  + m^2-q \right) G^3 = 0. \nonumber
\end{gather}
Hence we can fix the parameter $q$:
\begin{align}
    q=m^2+d-1,
\end{align}
and find the value of $(\xi \cdot\xi)$:
\begin{align}
    (\xi \cdot \xi) = 1 - m^2.
\end{align}
This means that, depending on the value of the parameter $m$, the vector $\xi$ can be timelike, lightlike, or spacelike. In Fig.~\ref{Glob de sitter 1} we show two cases, corresponding to lightlike and spacelike separation. We note that the lightlike case completely coincides with the one described in the previous paper \cite{Akhmedov:2026msi}, since in this case $m=1$ and the coefficients of $\sin(\phi/2)$ in the two theories coincide.

We also emphasize an important difference. For a non-null vector $\xi$, in higher dimensions we cannot set the soliton profile by an arbitrary function $F$ of ratios of hyperbolic waves with different null vectors, as was possible in the lightlike case. At the level of equations this is because, when $\xi$ is null, in the analogous equation \eqref{two line} both lines vanish simultaneously \cite{Akhmedov:2026msi}. After the replacement $G \to G F$, where $F$ is the soliton profile, all contributions did not mix and automatically vanished. In the new case this homogeneity is lost, and cancellation requires different lines, so the simple substitution $G \to GF$ no longer works. Moreover, this solution is interesting because it has no flat-space analog, since it becomes trivial in the flat limit:
\begin{align}
    \lim_{R\to \infty} \phi \to 2\pi.
\end{align}
We also note that this is still a single-soliton solution, and we have not yet been able to find multi-soliton solutions with either null or non-null vectors.

\subsection{Static solution and Energy}
Consider the Poincare coordinates:
\begin{align}
\label{Poincare}
\begin{cases}
X_{1}=R{\frac {z}{2}}\left(1+{\frac {1}{z^2}}\left(1+\frac{{x}^{2}-t^{2}}{R^2}\right)\right)
\\X_{2}={\frac {1}{z }}t
\\X_{i}={\frac {1}{z }}x_{i-2}  \quad \quad i \in (3,...,d+1)
\\X_{d+2}=R{\frac { z}{2}}\left(1-\frac{1}{z^2}\left(1-\frac{{x}^{2}-t^{2}}{R^2}\right)\right)\end{cases}.
\end{align}
To find a static solution, we must require that $(X \cdot \xi)$ does not depend on time; hence in the present case we set: 
\begin{align}
    \xi_1=\xi_{d+1}, \quad  \xi_2=0, \quad \vec\xi=\xi_i \ne 0, \quad i \in  (3,...,d+1)
\end{align}
such that:
\begin{align}
    (\xi \cdot \xi) = 1-m^2=(\vec{\xi}\cdot\vec{\xi})  >0.
\end{align}
Hence, for small mass $m^2<1$, the vector $\xi$ is spacelike and we obtain a static solution in dimensions greater than two:
\begin{align}
    \phi=4\arctan \left(\frac{\vec{x} \cdot \vec{\xi}-\xi_1}{z}\right).
\end{align}
The energy of such a solution diverges as:
\begin{align}
    E= \int_{0}^{\infty} \frac{dz}{z^d}\ldots
\end{align}
which is a common situation in AdS space due to the peculiar behavior of fields near the boundary of spacetime. Thus, here we have the same behaviour as in the null case \cite{Akhmedov:2026msi}, there the static solution also has infinite energy for $m<\frac{d+1}{2}$.

\section{Conclusion and acknowledgments}
We consider double sine-Gordon theories:
\begin{align}
    \Box \phi - m^2 \sin \phi - 2 q \sin \frac{\phi}{2} = 0,
\end{align}
whether parametr $q$ is not fixed. Below we present a summary in Table~\ref{table} of what we currently know about solutions for the two different parameters $q$.

\begin{table}[htbp]
\centering
\begin{tabular}{|c|c|c|}
\hline
 & $q=\frac{d m}{R}$ & $q=m^2+\frac{d-1}{R^2}$ \\
\hline
$\phi$ in $1+1$ dimensions & $4 \arctan\left[  \left(\frac{X \cdot \xi }{R}\right)^{m R}\right]$ & $4 \arctan\left[  \frac{X \cdot \xi }{R}\right]$ \\
\hline
$\phi$ in dimensions $>2$ & $4 \arctan\left[  \left(\frac{X \cdot \xi }{R}\right)^{m R}F(...)\right]$ & $4 \arctan\left[  \frac{X \cdot \xi }{R}\right]$ \\
\hline
$(\xi \cdot \xi)$ & 0 & $1-(m R)^2$ \\
\hline
$m R$  & $\in \mathbb{N}^+$ & $\in \mathbb{R}^+$ \\
\hline
Infinite energy for static solution & for $m R<\frac{d-1}{2}$ & for $mR<1$ \\
\hline
\end{tabular}
\caption{Properties of the solutions for the two values of the parameter $q$.}
\label{table}
\end{table}
One of our conjectures is that this class of theories can be embedded in the supersymmetric generalization of the sine-Gordon theory in $AdS$ spacetime, such that $q\sim \bar{\psi} \psi$ on such single solition solution. We hope that this approach can be generalized to obtain more general solutions that reduce to the general $n$-soliton sine-Gordon solution in the flat-space limit. However, we cannot exclude the possibility that multi-soliton solutions in hyperbolic spaces cannot exist in principle.

We would like to thank E. Akhmedov, K. Bazarov, K. Gubarev, K. Kazarnovskii, I. Myakutin, D. Sadekov, and G. Zverev for valuable discussions. The work of Dmitrii Diakonov was supported by the grant No. 26-12-00330 from the Russian Science Foundation (RSF).

\bibliographystyle{unsrturl}
\bibliography{bibliography.bib}

\end{document}